\documentclass[utf8]{frontiersSCNS} 

\usepackage{url,hyperref,lineno,microtype,subcaption}
\usepackage{lscape}
\usepackage{csquotes}
\usepackage[onehalfspacing]{setspace}
\linenumbers

\def\keyFont{\fontsize{8}{11}\helveticabold }
\def\firstAuthorLast{Turnbull {et~al.}} 
\def\Authors{Steven Martin Turnbull\,$^{1,2*}$ and Dion R.J. O'Neale\,$^{2,3}$ }
\def\Address{$^{1}$School of Critical Studies in Education, Faculty of Education and Social Work, University of Auckland, Auckland, New Zealand \\
$^{2}$ Te Punaha Matatini, University of Auckland, Auckland, New Zealand\\
$^{3}$ Department of Physics, University of Auckland, Auckland, New Zealand
}
\def\corrAuthor{Steven Martin Turnbull}

\def\corrEmail{s.turnbull@auckland.ac.nz}

\begin{document}

\nolinenumbers

\onecolumn
\firstpage{1}

\title[Entropy of STEM Co-Enrolment Networks]{Entropy of Co-Enrolment Networks Reveal Disparities in High School STEM Participation} 

\author[\firstAuthorLast ]{\Authors} 
\address{} 
\correspondance{} 

\extraAuth{}

\maketitle

\begin{abstract}

The current study uses a network analysis approach to explore the STEM pathways that students take through their final year of high school in Aotearoa New Zealand. By accessing individual-level microdata from New Zealand's Integrated Data Infrastructure, we are able to create a co-enrolment network comprised of all STEM assessment standards taken by students in New Zealand between 2010 and 2016. We explore the structure of this co-enrolment network though use of community detection and a novel measure of entropy. We then investigate how network structure differs across sub-populations based on students' sex, ethnicity, and the socio-economic-status (SES) of the high school they attended. Results show the structure of the STEM co-enrolment network differs across these sub-populations, and also changes over time. We find that, while female students were more likely to have been enrolled in life science standards, they were less well represented in physics, calculus, and vocational (e.g., agriculture, practical technology) standards. Our results also show that the enrolment patterns of the M\={a}ori and Pacific Islands sub-population had higher levels of entropy, an observation that may be explained by fewer enrolments in key science and mathematics standards. Through further investigation of this disparity, we find that ethnic group differences in entropy are moderated by high school SES, such that the difference in entropy between M\={a}ori and Pacific Islands students, and European and Asian students is even greater. We discuss these findings in the context of the New Zealand education system and policy changes that occurred between 2010 and 2016.

\tiny
 \keyFont{ \section{Keywords:} Network Analysis, STEM Education, Assessment, Enrolment, Entropy} 
\end{abstract}

\section{Introduction}
There is an increasing demand to understand the choices that students make when it comes to selecting courses in secondary school and further education. Obtaining a clear picture of the skills that students leave school with is an important goal for governments across the world, and this is especially true regarding Science, Technology, Engineering and Mathematics (STEM). For example, the New Zealand Qualifications Authority (NZQA) \citep[p.8]{NZQA2016} specifically stated that:

\begin{quote}
    To meet the demand for essential skills for the twenty first century, New Zealand needs to grow the number and diversity of skilled workers in Science, Technology, Engineering and Maths.
\end{quote}  

Governments are pushing to not only increase the number of students participating in STEM education, but also to increase the representation of students who have been historically underrepresented in STEM. While trends may differ across countries, disparities in STEM participation tend to be found at the intersection of gender, ethnicity and social class \citep{Archer2015b, PISA_NZ_2017}. Globally, women are typically underrepresented in subjects such as physics and computer science, while there tends to be gender parity in subjects such as biology and medicine. In the case of Aotearoa New Zealand, similar disparities in STEM participation are found \citep{NZQA2016,EducationCounts_2016a,EducationCounts_2016b}. In addition, students from M\={a}ori and Pacific Island backgrounds have also been underrepresented in post-compulsory STEM education \citep{MoH2014, NZQA2016}. 

Student attrition from STEM education is often viewed in terms of a leaky pipeline, with students from groups who are under represented in STEM being more likely to drop out of STEM education with each advance from one educational stage to the next. However, participation in STEM education is complex. Not only is it important to consider the socio-cultural context in which students are placed when they make their subject choices, it is also important to consider the structural context of the education system. We are increasingly able to draw upon rich, complex, education-related administrative data to achieve this, but we must consider how we can analyse these data in a manner that preserves complex structures and provides new and useful insights. By meeting this goal, we can increase our understand of what participation in STEM looks like.

As detailed by \citet{hipkins2005staying}, there are many ways in which STEM participation can be reported on. At a broad level, we can summarise the number of students enrolled in each subject (e.g., how many students of each demographic group study biology?). We can also explore patterns at finer-grained levels by summarising participation per high school (e.g., which high schools have higher proportions of students studying science?), or by reporting participation at the level of assessment (e.g., how many students took a specific biology exam?). While it is relatively easy to summarise and interpret participation at broad levels, untangling and understanding patterns of subject participation at fine-grained levels can be a difficult task. This task is especially difficult in the context of Aotearoa New Zealand, which operates a particularly complex high school assessment system. 

The goal of the current study is to develop and employ a novel method of reporting on student participation in STEM by looking specifically at students' co-enrolments at the level of assessment. We begin by summarising the insights that can be gained by exploring STEM participation at a broad level. We then provide a brief summary of the National Certificate of Educational Achievement (NCEA), Aotearoa New Zealand's internationally unique high school qualification. We then move on to demonstrate how the quantitative technique of network analysis can be employed to reveal structures in NCEA participation. Finally, we discuss the novel insights provided by network analysis of STEM co-enrolments in NCEA assessments spanning the previous decade. \\

\subsection{Broad Understandings of Student Participation in STEM}
Student participation in STEM is often reported at a broad level, with information detailing the counts of students who are enrolled in each subject, and how this differs across demographic groups. In Aotearoa New Zealand, data is readily available by sex (male or female) and Socio-economic status (SES) from 2004 to 2018 \citep{EducationCounts_2018}. Exploring these data can provide a surface level description of what the field of STEM education looks like in Aotearoa New Zealand.

\begin{figure}[htpb]
    \centering
    \includegraphics[width = \textwidth]{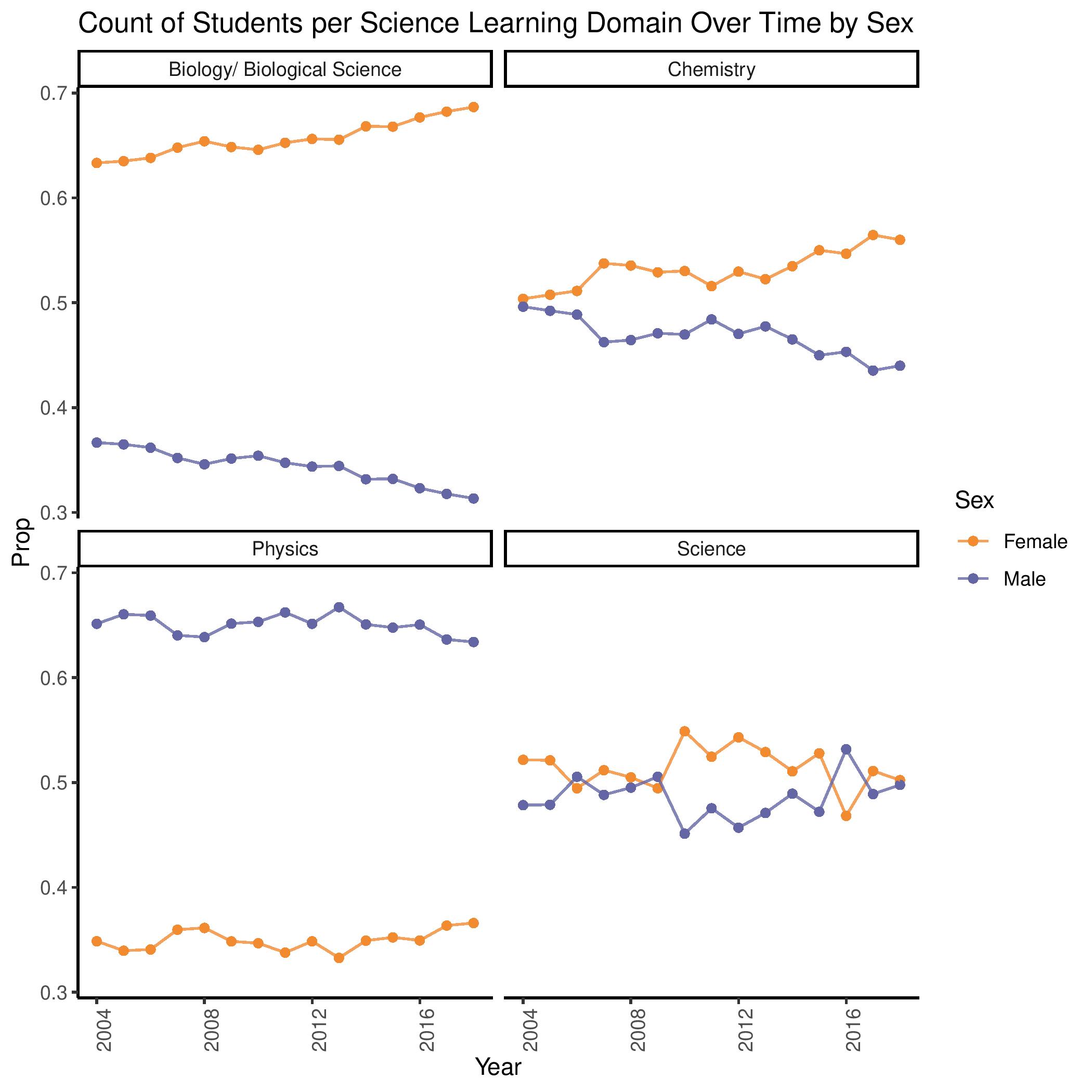}
    \caption{\textbf{Science Participation Rates Across Years by Sex}. These plots show the participation of male and female students in key science subjects in Year 13 from 2004-2018. Biology and Chemistry had a greater share of female students (nearly 70\% of biology students in 2018 were female). Physics continues to be male-dominated. ``Science'' represents core science assessments, which assess topics about more general aspects of science (including applications to everyday life and societal issues). This core science domain had a relatively balanced representation of male and female students across years. Data retrieved from \citet{EducationCounts_2018}.}
    \label{fig:STEM_Participation_Sex}
\end{figure}

As shown in Figure \ref{fig:STEM_Participation_Sex}, in Year 13 (final year of high school), female students in Aotearoa New Zealand are less likely to take physics, with this under-representation being steady across years. The same figure shows that female students are more likely to take biology, and more recently chemistry, with this over-representation becoming increasingly more pronounced over time.

\begin{figure}
    \centering
    \includegraphics[width = \textwidth]{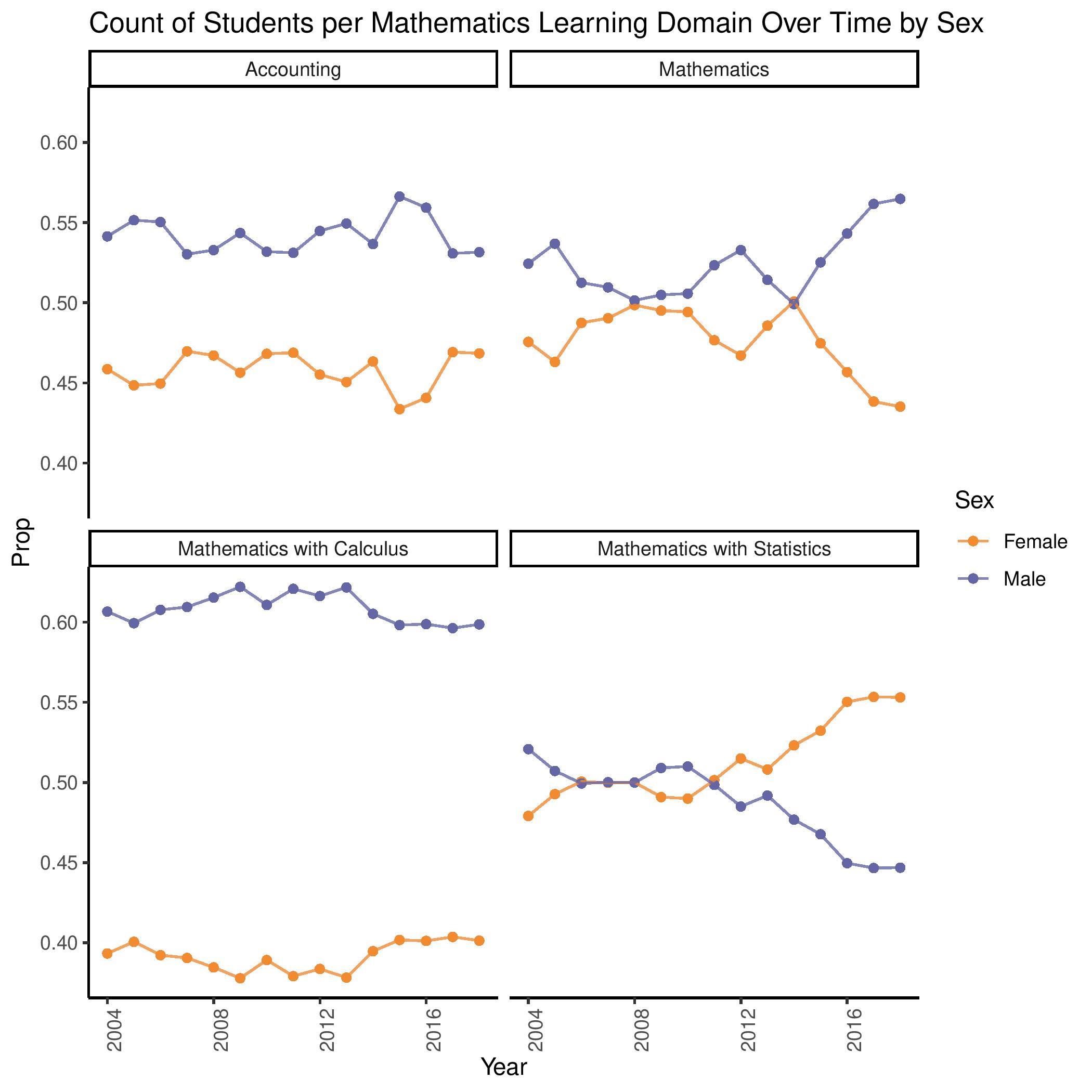}
    \caption{\textbf{Mathematics Participation Rates Across Years by Sex}. These plots show the participation of male and female students in key mathematics subjects in Year 13 from 2004-2018. There have been relatively even levels of participation in mathematics subjects over time, except calculus where male students continue to be enrolled in greater numbers. Data retrieved from \citet{EducationCounts_2018}.}
    \label{fig:STEM_Participation_Sex_Maths}
\end{figure}

Figure \ref{fig:STEM_Participation_Sex_Maths} shows that female students continue to be underrepresented in mathematics subjects, such as accounting and calculus. However, female students have higher levels of representation in statistics than male students in recent years \citep{EducationCounts_2018}. The same data shows that, in technology subjects, the computer and engineering subjects have continually been male-dominated, with this becoming more pronounced over time\citep{EducationCounts_2018}. Food technology and textiles are the only female dominated technology domains. 

\begin{figure}
    \centering
    \includegraphics[width = \textwidth]{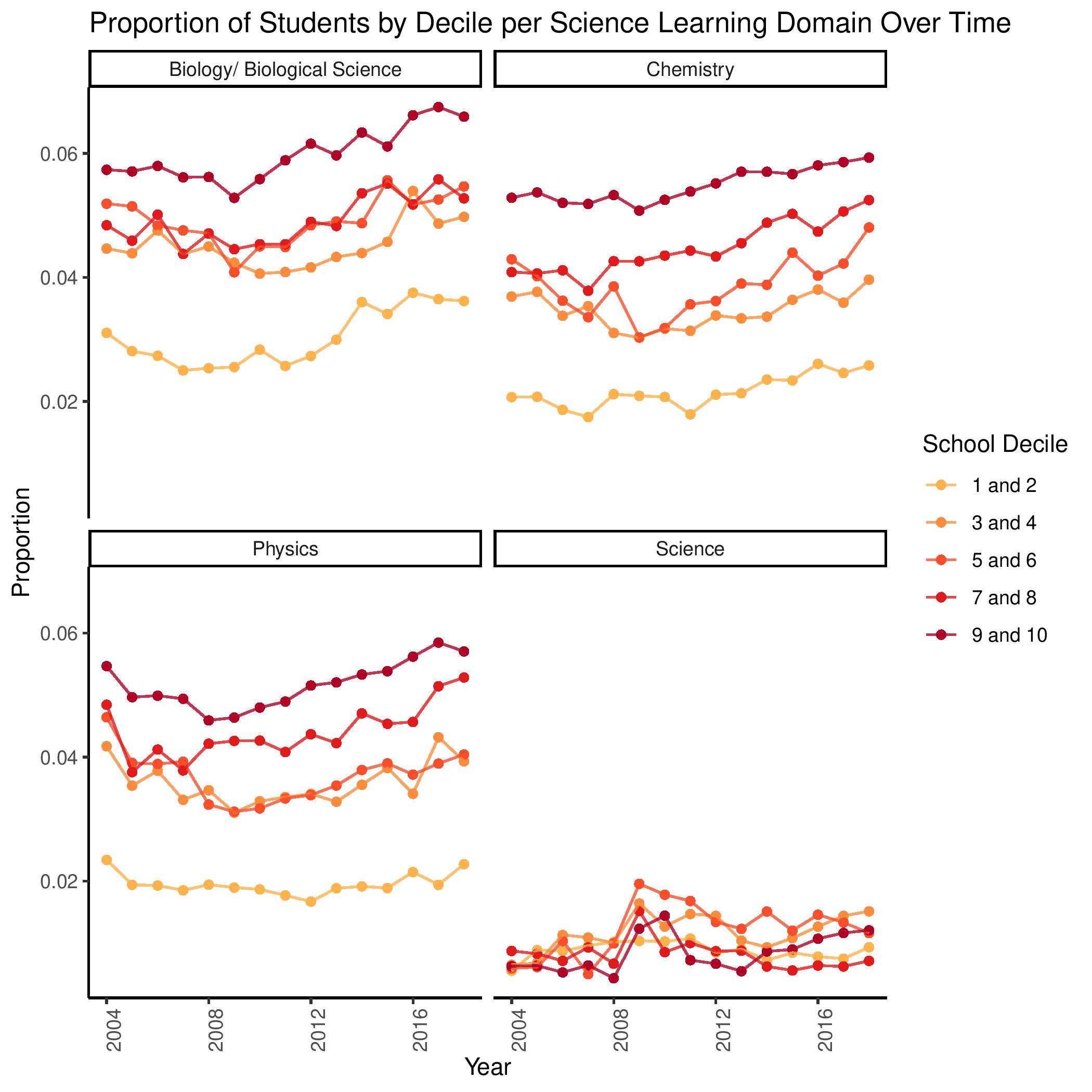}
    \caption{\textbf{Science Participation Rates Across Years by School Decile}. School deciles are grouped into quintiles (deciles 1 and 2 together, deciles 3 and 4, and so on). These plots show the rate of participation for each decile group as a function of the total subject enrolments (across all learning domains) for that group. This takes into account that higher decile groups contain a greater number of students than lower deciles. As can be seen in the above plots, students from higher decile schools are more likely to take biology, chemistry, and physics. Low decile schools had a flatter participation rate in physics, but increasing participation over time in biology and chemistry, such that the increase is similar to higher decile schools. The plot of the core ``Science'' subject appears less ordered and more variable in terms of decile ordering, but also has a lower rate of participation overall. Data retrieved from \cite{EducationCounts_2018}.}
    \label{fig:STEM_Participation_Decile}
\end{figure}

Data from \cite{EducationCounts_2018} also allows us to see trends in STEM participation by school decile, a proxy measure of SES. In Aotearoa New Zealand, school decile refers to the affluence of the neighborhood in which a school is located. High decile schools are located in more affluent areas, whilst low decile schools are located in less affluent areas. As shown in Figure \ref{fig:STEM_Participation_Decile}, students  who attended higher decile schools had greater rates of participation in science subjects, and this pattern was also evident for calculus and statistics. The relationship between student enrolment in technology learning domains and decile has no discernible pattern.  

Broad level data, such as those discussed above, allow us to interpret trends in subject enrolments over time. However, they provide only a surface level understanding of STEM participation. Beneath the aggregation of counts per subject label hides important information that is useful for policy makers and researchers. Each subject consists of many different assessments, each covering unique content and following different assessment criteria. The following section will provide a brief introduction to Aotearoa New Zealand's main high school assessment system, the National Certificate of Educational Achievement (NCEA).\\

\subsection{A Brief Introduction to the National Certificate of Educational Achievement}
The National Certificate of Educational Achievement (NCEA) is the main form of secondary school assessment in Aotearoa New Zealand. First introduced to students in 2002, NCEA was designed to replace norm-referenced assessment. In norm-referenced assessments student achievement is judged against the average achievement of the student population \citep{Mahoney2005}. Instead, achievement in NCEA is based on the competencies of individual students \citep{hipkins2016ncea}, meaning that achievement is an indicator of what a student \textit{knows}, and not just how they rank amongst their peers. Therefore, it is possible for all students to pass if they all meet the assessment criteria. Assessment operates at the level of specific skills, or \textit{standards}, that comprise a subject discipline. For example, instead of just receiving an overall grade for biology, students take several standards in the subject discipline of biology that demonstrate their competence in particular areas (e.g., ``\textit{Demonstrate understanding of biological ideas relating to micro-organisms}''). By successfully completing standards, students accumulate credits, the value of which is linked to the amount of work needed to fulfil a standard. The three levels of the NCEA typically correspond to the final three years of high school. NCEA Level 1 is typically taken in Year 11 (age $\sim$ 15), NCEA Level 2 in Year 12, and NCEA Level 3 in Year 13.

What makes the NCEA a unique assessment system is its flexibility. Compared to the systems it replaced (School Certificate, Sixth Form Certificate, and Bursary), there is more variety in the assessments/standards that students may be enrolled in \citep{Mahoney2005}. In providing increased choice to students and their educators, and more flexible pathways through high school, it was hoped that the NCEA would benefit students from a range of backgrounds. As stated in the New Zealand curriculum \citep[p.41]{NZCurriculum2007}: \begin{quote}
Schools recognise and provide for the diverse abilities and aspirations of their senior students in ways that enable them to appreciate and keep open a range of options for future study and work. Students can specialise within learning areas or take courses across or outside learning areas, depending on the choices that their schools are able to offer.
\end{quote}
The NCEA meets these goals by providing students with more learning pathways through high school, which aims to serve both students who wish to progress to tertiary study, and those who want to enter into vocational careers. These two pathways are reflected in the two main types of assessment offered: unit and achievement standards.\\

\subsubsection{Unit and Achievement Standards}
Unit standards tend to assess more vocational subjects (e.g., plumbing, hairdressing, agriculture). Unit standards have strict criteria that need to be achieved in order to pass \citep{hipkins2016ncea}, and are thus suited to assessing skills that follow a procedure. If a student meets the criteria they pass; if they fail a step, they fail the standard. All unit standards are assessed internally by the institution where the student is placed, offering the opportunity to teach and learn in a manner that caters more to students' contexts. Internal assessments are moderated by the NZQA, according to the New Zealand Qualification Framework\citep{NZQF}, to ensure the assessment is consistent and rigorous. That being said, schools often provide the opportunity for students to retake failed internal assessments at a later time.

Achievement standards assess more traditional subjects that are tied to the New Zealand curriculum, such as science, mathematics, and English. While many achievement standards are assessed internally, a number of them are taken under standardised conditions and assessed by an external body (i.e., the NZQA). Unlike unit standards, where students can only be judged to have passed or failed, achievement standards often have assessment criteria that can be interpreted more subjectively and require a different grading structure \citep{hipkins2016ncea}. Instead of pass or fail, achievement standards have four outcomes: not achieved, achieved, merit, and excellence. This grading structure seeks to reward students who demonstrate knowledge at a higher level than simply showing competence. The introduction of different grading levels in achievement standards provides increased opportunity to rank students by performance \cite{shulruf2010new}, a process that NCEA was not initially designed to accommodate \citep{hipkins2016ncea}.

The relevance of achievement and unit standards can be tied to students' future aspirations in the context of STEM. \citet[p.20]{wong2016science} differentiates these aspirations as being tied to either careers \textit{in} science, or careers \textit{from} science. Careers \textit{in} science may be defined as: ``...occupations that are involved with the research or discovery of science as their primary purpose''  \citep[p.20]{wong2016science}. Achievement standards may be more closely linked to these types of careers as they provide the means to assess theoretical work, and provide the pathway to university. Careers \textit{from} science may be defined as ``careers that are related to science'' but prioritise other aspects of STEM  \cite[p.20]{wong2016science}. This includes careers in technology, and also careers in horticulture and farming that are even more applied. The vocational slant of unit standards may prepare students better for these types of careers \textit{from} science. With that being said, students can take a combination of unit and achievement standards.

While there are many potential pathways through the NCEA, the eventual goal for students is to accumulate enough credits to achieve NCEA Level 3. Students who wish to attend university must meet a separate goal over and above the requirements for NCEA Level 3. To be eligible to enrol at a university, students must attain University Entrance (UE), which is the equivalent of achieving NCEA Level 3 with a specified number of credits coming from three subjects on an approved subjects list (these include subjects such as biology, physics, mathematics, and English) with specific achievement standards \citep{NZQA_Approved}, and a higher standard of literacy than regular NCEA Level 3 \citep{hipkins2016ncea}. Specific university programs may also have their own requirements for enrolment. For example, to transition from NCEA to engineering at the University of Auckland, students must attain specific externally assessed achievement standards in Level 3 calculus and physics \citep{UoA_Engineering}. Alternate pathways to university STEM study are possible, such as completion of university foundation courses, but these take additional time. The decisions that students make regarding the selection of STEM standards in NCEA Level 3 can thus have long-lasting implications. It is therefore especially important to understand how NCEA Level 3 is structured, and how this relates to student outcomes.  

Given the complexity of the NCEA, exploring participation in STEM at the level of individual assessments can provide additional insights that complement our broad level understandings of STEM participation discussed previously. Doing so allows us to explore factors related to individual standards (such as the type of standard assessment and whether it was assessed internally or externally) as well as co-enrolment patterns and pathways through assessments. To build on the broad level understandings outlined above, we now adopt the following research questions:
\begin{itemize}
    \item Can we identify patterns in the NCEA Level 3 standards taken by students in STEM?
    \item If so, how do the patterns of NCEA Level 3 standard enrolments differ across demographic characteristics, SES, and time?
\end{itemize}

Given that the NCEA can be considered ``one of the most complicated education system in the world'' \citep{hipkins2016ncea}, unpacking details at a more fine-grained level can be a daunting task. To explore this complicated system and answer our research questions, we employ quantitative techniques based in the field of network analysis. We explain how network analysis can be used as a tool to understand patterns of assessment, especially in contexts where the system is complex (as with the NCEA). The following sections will discuss how network analysis can help us explore what participation looks like for students studying STEM. \\

\section{A Network Methodology}
\subsection{Data}
We make use of Statistics NZ's Integrated Data Infrastructure (IDI) to access administrative data pertaining to students' high school and census information \citep{IDI}. The IDI is a collection of government data sets, containing micro-data on student enrolment and demographics, linked at the level of individuals for the population of Aotearoa New Zealand. We focus on students taking NCEA Level 3 from 2010 to 2016, as this is the most up to date data available at the time of writing. We focus on NCEA Level 3 as this level is the most highly specialised, and precedes entrance to university and employment. Years prior to 2010 are available, but were omitted due to processing constraints. The years spanning 2010 to 2016 were also of specific interest, due to education policy reforms introduced around 2012 and 2013 \citep{hipkins2016ncea}. 

We apply several rules when selecting student cohorts to be included, in order to minimise the risk of adding statistical noise to our analysis. In order to focus our analysis on students who have had the majority of their education in Aotearoa New Zealand, we only select individuals who are identified as having tax, birth or visa records present in the IDI. We also only include students who had NCEA records when they were 15 or 16 and during NCEA Level 1. These filters help focus our sample on the resident population of Aotearoa New Zealand, and minimise the chances of including visitors or foreign exchange students. We also limit our sample to students who attended state schools in Aotearoa New Zealand. This is because private schools in Aotearoa New Zealand are more likely to offer a combination of the NCEA and other formal qualifications (such as Cambridge or International Baccalaureate), introducing additional layers of complexity. For the purposes of our analysis we also assign each student a single cohort year based on the most frequent year in which they took standards. This is because students are able to take NCEA Level 3 standards over multiple years. For example, if a student took two NCEA Level 3 standards during 2015, and ten NCEA Level 3 standards during 2016, we would assign the student to the 2016 cohort. We choose not to exclude Level 3 standards taken in a different year from the overall cohort year, as these standards would still contribute to the student's qualification.

We include the following variables in our analysis:
\begin{itemize}
    \item Students' sex (male or female). Due to limitations in the administrative data used, we are not able to include gender (and non-binary classifications of gender) in our analysis.
    \item Students' ethnicity. Each student is able to identify with multiple ethnic groups, following the classification set out by Stats NZ \citep{StatsEthnicity}. The main ethnic groups include European; M\={a}ori; Pacific Island; Asian; Middle Eastern, Latin American, or African (MELAA); and Other. For the purposes of this study, we do not report results for MELAA and Other populations as they include a broad cross section of individuals, but typically involve relatively small numbers.
    \item High school decile. This is a rating out of 10 for the affluence of the area where the school is located. For the purposes of the following analysis, we categorise high school decile into 3 groups. Deciles 1-3 are low decile, deciles 4-7 are medium decile, and deciles 8-10 are high decile. 
    \item NCEA Level 3 standards taken. For each student, we have records of all of the standards taken at NCEA Level 3. We only include standards from the New Zealand curriculum learning areas of Science, Technology and Mathematics \citep{NZCurriculum2007}. For each standard, we have information on its subject area (e.g., physics, biology, mathematics etc.), whether it was a unit or achievement standard, and whether it was assessed internally or externally.\\
\end{itemize}

\subsection{Network Analysis}
We employ network analysis to understand STEM enrolment at NCEA Level 3 at a fine-grained level. At its fundamental level, a network is a collection of nodes and edges. Nodes can represent an agent (e.g., a student) or an object (e.g., a standard); edges link two nodes together to indicate some form of relationship. Networks can be used to represent anything from human relationships and transport networks, to biological and computer systems \citep{barabasi2003linked}. In education research, network analysis has tended to focus on the relationships shared between students in the classroom \citep{tranmer2014multiple}, or communication between staff at educational institutions \citep{daly2010social}. There are few examples of education research that use network analysis to investigate non-social relationships. We seek to expand this area of research by applying network analysis to high school assessment enrolment data. As we will outline in the following section, network analysis can help us identify patterns in NCEA standard co-enrolments.

\begin{figure}
    \centering
    \includegraphics[width = \textwidth]{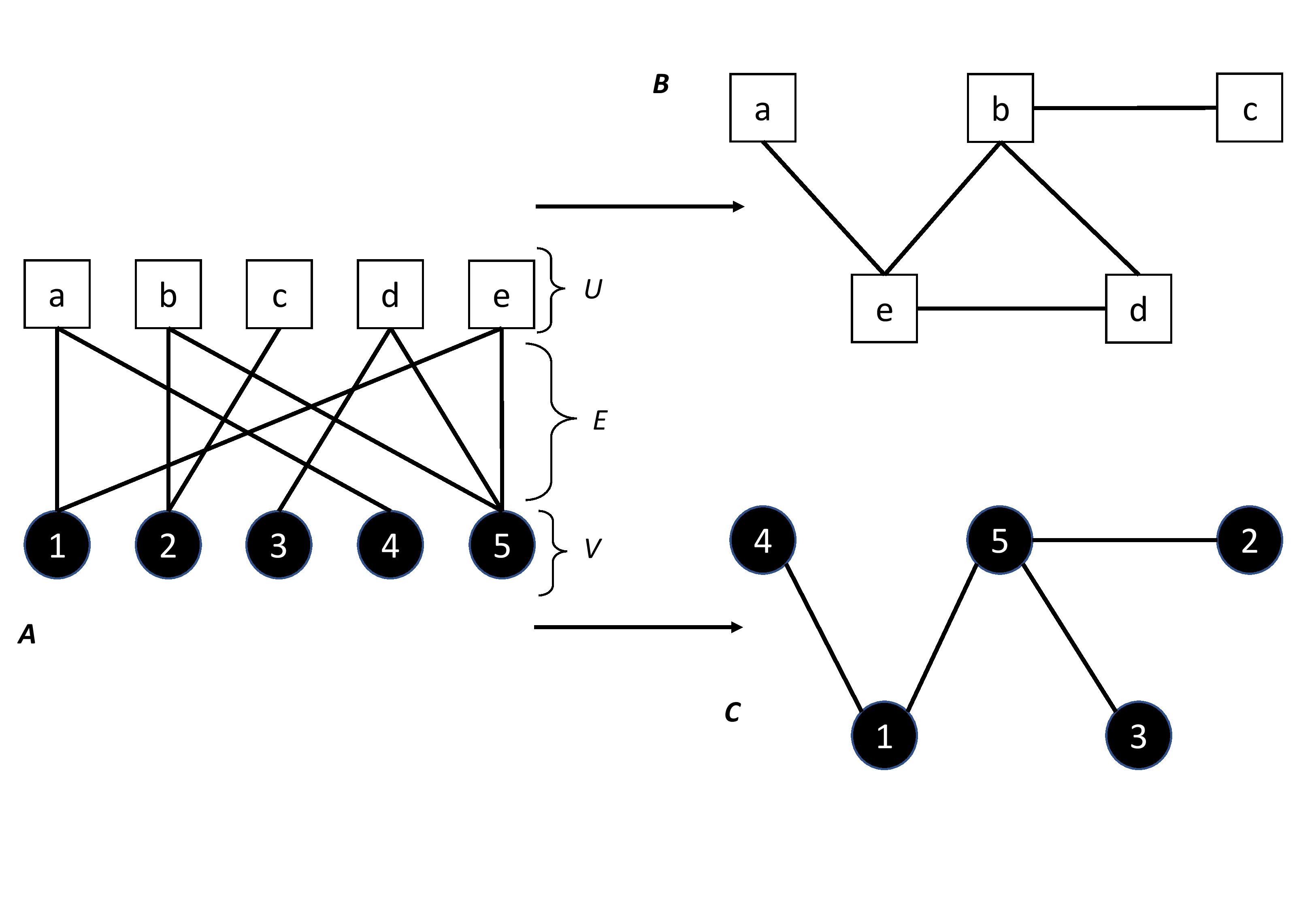}
    \caption{\textbf{An Example of a Bipartite Network and its Projections}. In the case of the current study, white nodes (set \textit{U}) represent standards, and black nodes (set \textit{V}) represent students. A) Nodes of different sets are connected by an edge \textit{E} (i.e., an edge will exist if a student took a particular standard). B) The projection of set \textit{U}. In the current study, we use this projection to represent a network of standards. Two standards will be connected by edge if a student enrolled in both standards. C) The projection of set \textit{V}. In the current study, this refers to a network of students, with edges indicating that students both took the same standards. To help preserve students' confidentiality, we do not report on this projection.}
    \label{fig:BipariteNetwork}
\end{figure}

In our analysis, nodes take the form of students and standards. Edges in our network represent any recorded instance where a student was enrolled in a NCEA Level 3 STEM standard during high school. This creates a bipartite network (also commonly referred to as a two-mode network). A bipartite network is any network where there are two types of node, and nodes can only connect to a node of a different type. In our case, a standard cannot be connected directly to another standard, and a student cannot be connected to another student. For example, in A in Figure \ref{fig:BipariteNetwork}, standards may be represented by nodes in set \textit{U}, and students may be represented by nodes in set \textit{V}.

We create a network of students and the standards they were enrolled in for the whole of our student population. We structure this network so that it is multidimensional. Each student node belongs to a specific year, region, and decile, while standards can exist across multiple years, regions, and school deciles. In order to analyse the properties of our network, we are required to `project' onto one set of nodes. This means that we take the node set belonging to a single node type, and generate edges between these nodes when they are linked to a common node of the other node set. For example, B and C in Figure \ref{fig:BipariteNetwork} shows the projection of the network in Figure \ref{fig:BipariteNetwork}. In the projections, standards represented in set \textit{U} are now connected to one another (B), and students in set \textit{V} are also now connected (C). 

As we are interested in the patterns of standards that students took, we project onto the standard nodes (Figure \ref{fig:BipariteNetwork}:B). This results in a network of standards that are connected by edges indicating that students took those two standards together within their NCEA Level qualification. The edges of the projected standard network can also take on a weighting that corresponds to the frequency that two standards were taken together by students.\\ 

\subsection{Normalization and Community Detection}
Our goal is to use the co-enrolment network to understand the standards that tend to be taken together, and by which students. To do this, we employ community detection. Community detection is a process in which we identify sets of nodes that are clustered together by the edges in the network. Previous research by \citet{ferral2005clustering} has employed similar clustering techniques to investigate communities of subjects that tend to be taken together in the NCEA, but this was limited by the number of high schools sampled, the response rate of schools, and the availability of demographic and standard information. Usually, community detection methods identify communities by maximising the \textit{modularity} score within communities. Modularity refers to the tendency of nodes to connect to other nodes within the same community relative to nodes that are outside the community. While there are many different community detection algorithms, the current study makes use of the infomap algorithm \citep{rosvall2009map}. 

In order for our communities to more truly represent the standards that tend to be taken together, we need to normalize our edges so that weights do not refer to the raw counts of students' co-enrolments. The raw weighting does not consider the fact that standards have different populations of students. As a result, community detection may group two standards together simply because one standard has a large number of students. To explain more clearly, we can use the hypothetical case of English Standard A, Physics Standard A and Physics Standard B. If English Standard A has a population of 1000 students, and 10\% of those students take Physics Standard B, the raw weight is 100 students. If 100 students took Physics Standard A, and 90\% of those students took Physics Standard B, the raw weight is 90 students. While we would expect the two physics standards to be grouped together, using raw counts of students as edge weights may not result in grouping that meet these expectations. Instead, we make use of a normalization technique called Revealed Comparative Preference (RCP) to provide more consistent communities. RCP measures the fraction of students from standard $j$ who also took a second standard $i$, relative to the overall fraction of students taking standard $i$, across all other standards. More specifically: 
$$RCP(i,j) = \frac{x_{ij}/\sum\limits_{j}x_{ij}}{\sum\limits_{i}x_{ij}/\sum\limits_{ij}x_{ij}}$$
where $x_{i,j}$ is the number of students taking both standard $i$ and $j$, $x_j$ ( or $x_i$) is the total number of students taking standard $j$ (respectively, standard $i$), and $x$ is the total number of unique students enrolled in any standard. This RCP metric is based on the measure Revealed Comparative Advantage, used in economics \citep{Balassa1965}, and was calculated using the EconGeog package \citep{balland2017economic} in R \citep{team2013r}. The RCP calculation returns a value where anything greater than 1 indicates a `preference' for two standards being taken together. A value below 1 indicates that, given the number of students in either standard, there was no preference for the two standards being taken together. 

We remove any edge in the network where the RCP value is below 1, and subsequently any node that no longer has any edges (isolated nodes with a degree of 0). This results in a network that consists of standards connected by edges with a weighting relative to the preference for each standard being taken together with its neighbors in NCEA Level 3. We then identify communities of standards that are grouped together in our network using the infomap community detection algorithm \citep{rosvall2009map}. In simple terms, the infomap algorithm partitions the network in a way that maximizes the number of edges within a community, relative to the edges between communities. \\

\subsection{Exploring Participation}
To compare student participation across the educational fields detected, we can consider the relative proportion of students from particular years, and across school deciles and social groups. One of our goals is to establish an idea of how each network is structured. Are the enrolments for a specified demographic group spread more evenly across a network, or are they focused in particular areas? To answer this question, we employ the metric of entropy. Entropy is a concept originating from the field of thermodynamics, and provides an indication of how organised or disorganised a system is. In the case of the current study, we use entropy to assess how participation is spread across the network. Using the measures of entropy as signals of disparities, we then explore the rates of participation across communities and standards in finer detail. The following section will outline our measure of entropy.\\

\subsubsection{Entropy}
Entropy provides an aggregated metric of how ordered a system is. Systems that are highly ordered have a lower level of entropy, while disordered systems have higher entropy. To use an analogy, a crystalline solid with atoms focused together on a regular grid has low entropy, while a gas with atoms randomly spread across a grid has higher entropy. Following this analogy, we may explain low entropy as an indication that a pattern of standard enrolments is more focused or specialised in specific areas. In contrast, high entropy in the network of standards indicates that a patterns of enrolment is more diverse. By partitioning our network into different social groups (e.g., across sex, ethnicity, and school decile) we can explore similarities and differences in network structures. 

We calculate entropy in two steps. Firstly, we work out the probability of a sub-population enrolling in a specific standard given the total number of enrolments in the network for that sub-population. This probability is given by:
$$p^q_i = \frac{\sum\limits_j x_{ij}^q}{{\sum\limits_{ij} x_{ij}^q}}$$
where $x^q_i$ is the number of students in a sub-population~$q$ enrolled in standard~$~i$, and $x^q$ is the total number of enrolments for that sub-population. Using this measure of probability, we calculate entropy using the following formula: 
$$S^q = -\sum_{i}^{N}{
\frac{p^q_i\log{p^q_i}}{\log{x^q}}
}$$

Where $p^q_i$ is the probability of a student from sub-population~$q$ enrolling in standard~$i$, given the overall total number of enrolments for that sub-population($x^q$). The resulting score $S^q$ provides an single positive value that indicates the entropy in the network, where a lower value indicates lower entropy (i.e., ordered patterns of enrolments), and a higher value indicates higher entropy ( i.e., more disordered enrolments). While we choose to normalise by the total number of enrolments for a sub-population, alternative methods of normalisation are also possible. For example we could normalise by the number of standards in the network. This would be equivalent to assuming that the probability of a student enrolling in a specific standard is independent of the standard. This assumption does not hold for two reasons. Firstly, standards have very different numbers of student enrolments. Secondly, different student groups are differently represented in different standards. Normalising by $x^q$ accounts for the number of students from a sub-population enrolled in a specific standard. However, it does not let us distinguish between effects due to the size/popularity of a standard and those due to differing preferences of specific populations for specific standards. A downfall of this approach is that it does not account for standards that have no enrolments for students in a specific sub-population.

We ascertain a level of confidence by using a bootstrapping method, where we vary the count $x^q_i$ in each standard $i$ by a uniform random amount of up to $\pm20\%$, and recalculate entropy. We repeat this process 1000 times for each entropy measure. \\

\subsubsection{Trends}
Following the entropy measure, we investigate how participation differs across demographic groups per standard by comparing raw counts, proportions, and probabilities. The communities identified provide a good indication of the standards that tend to be taken together, which allows us to explore rates of participation across groups of standards as well as individual standards. We are able to explore a range of attributes, such as such as the probabilities of sub-populations enrolling in a standard, with respect to sex, ethnicity, school decile, and type of standard (achievement/unit standards, internally/eternally assessed).  

Following the identification of different communities of standards, our goal is to explore the student enrolment patterns in these communities. Based on trends outlined previously by \citep{hipkins2016ncea} and based on data from \citet{EducationCounts_2018}, we make the following hypotheses:
\begin{itemize}
    \item Female students will be more likely to have enrolled in standards in communities related to biology.
    \item Male students will be more likely to have enrolled in standards in communities related to physics, calculus and computer science.
    \item Students who attended high decile schools will be more likely to have enrolled in externally assessed standards.
\end{itemize}
Less research has investigated the relationship between assessment type (achievement or unit) and STEM enrolment, but we may expect that students groups who historically succeed in traditional forms of education (high SES, European and Asian students) to be more likely to have enrolled in externally assessed achievement standards. Student groups who have historically been under-served by traditional assessment may be more likely to have enrolled in unit standards. \\

\section{Results and Discussion}
\begin{landscape}
\begin{figure}[htbp]
    \centering
    \includegraphics[width =0.9 \textwidth]{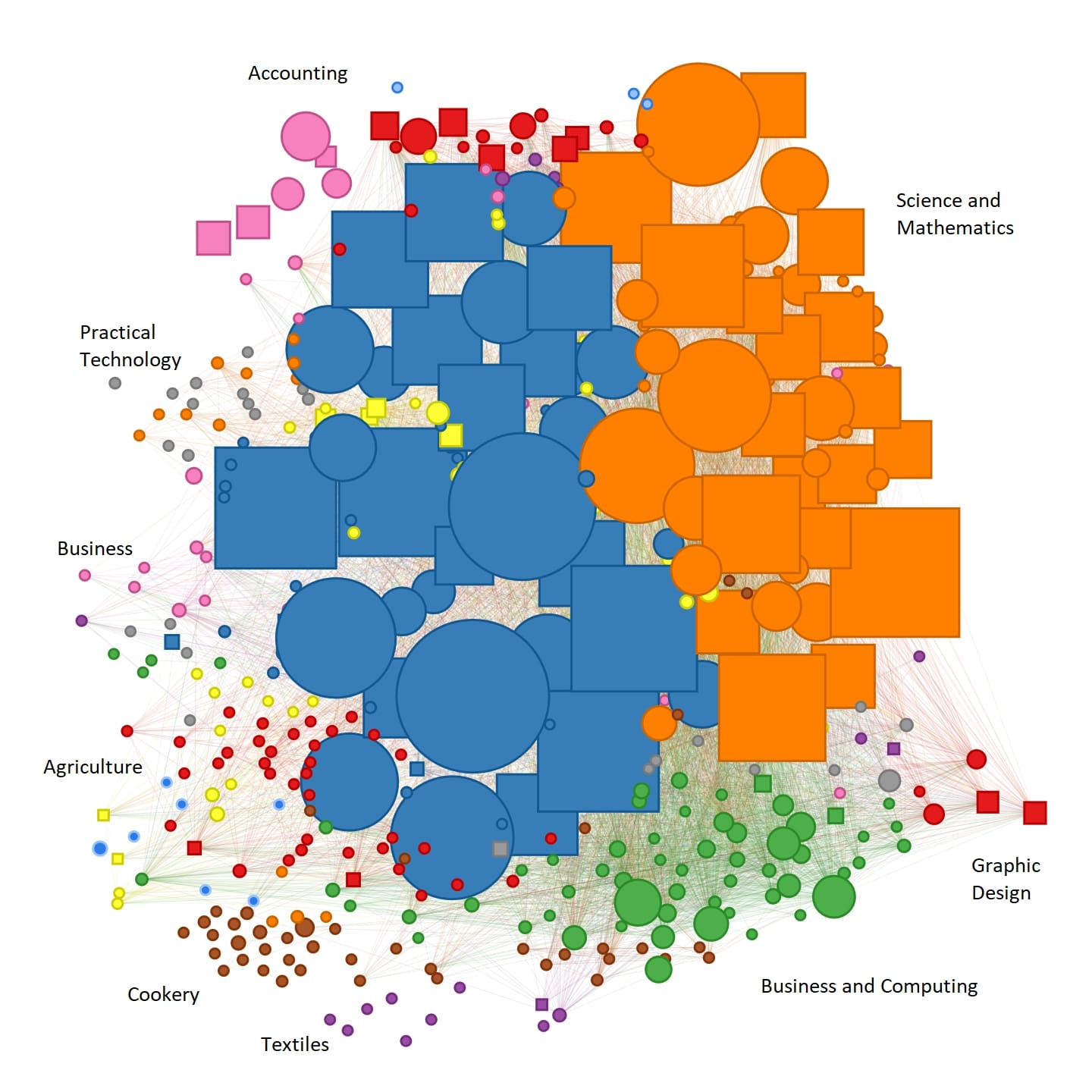}
    \caption{\textbf{Network of NCEA Level 3 Standards}. The standard projection of the NCEA Level 3 standard co-enrolment network. Nodes represent standards and edges represent a preference for two standards being enrolled in at the same time by students. Colours represent the communities of standards that tend to be taken together. Squared nodes represent externally assessed standards, and circular nodes represent internally assessed standards. The above network includes all standards offered between 2010 and 2016.}
    \label{fig:NetworkAll}
\end{figure}
\end{landscape}

\begin{landscape}
\begin{figure}
    \centering
    \includegraphics[width = 1.35\textwidth]{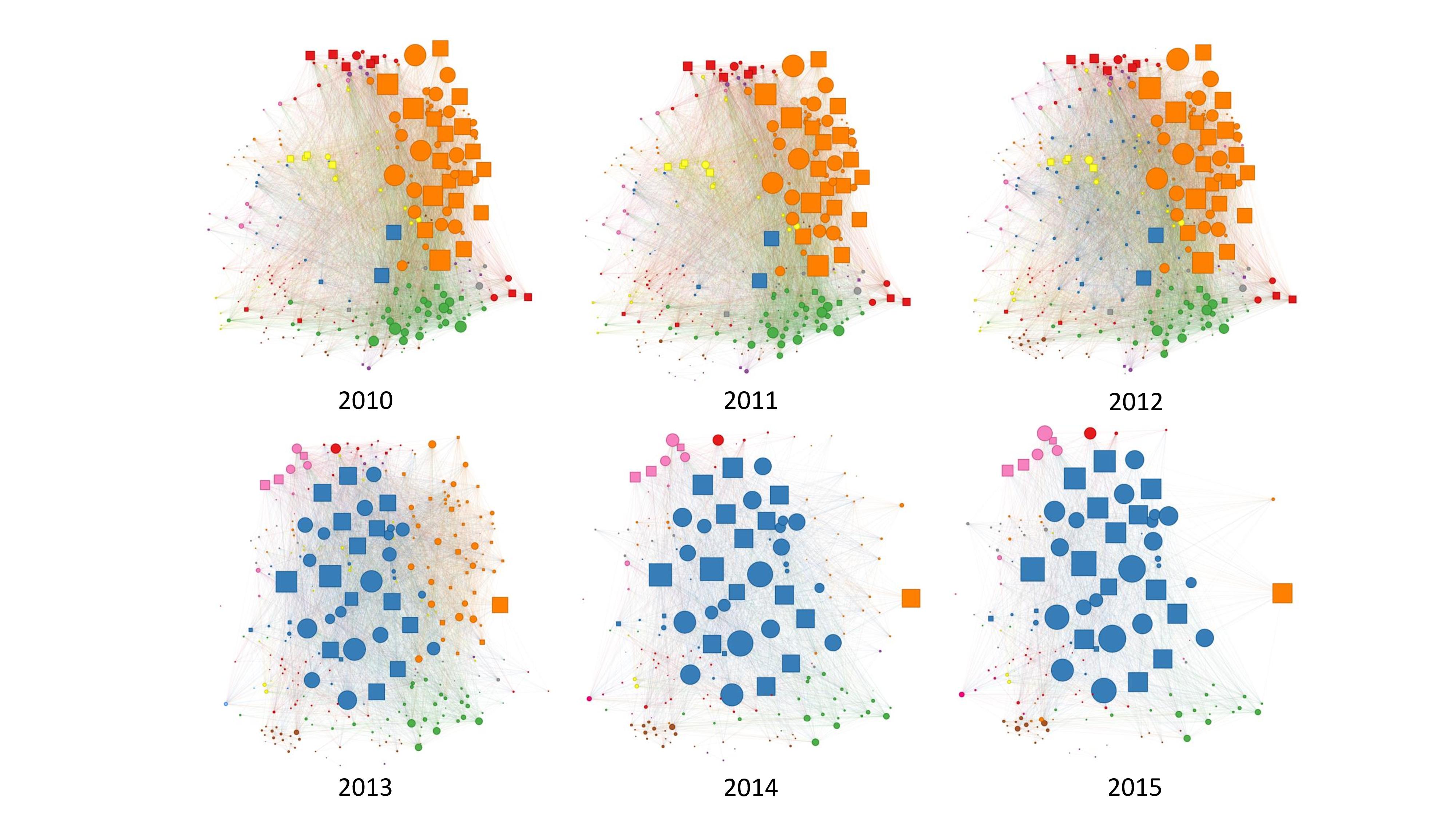}
    \caption{\textbf{Network of NCEA Level 3 Standards Across Years}. The standard projection of the NCEA Level 3 co-enrolment network across years. Node size represents the relative proportion of students enrolled in each standard. As years progressed, the number of standards was reduced. Around 2013, a new set of science and mathematics achievement standards were introduced. Post-2013, the network is comprised mainly of one mathematics and science community, which indicates that assessment was more standardised than previous years.
    }
    
    \label{fig:NetworkYear}
\end{figure}
\end{landscape}

\begin{figure}[htbp]
    \centering
    \includegraphics[width = 0.75\textwidth]{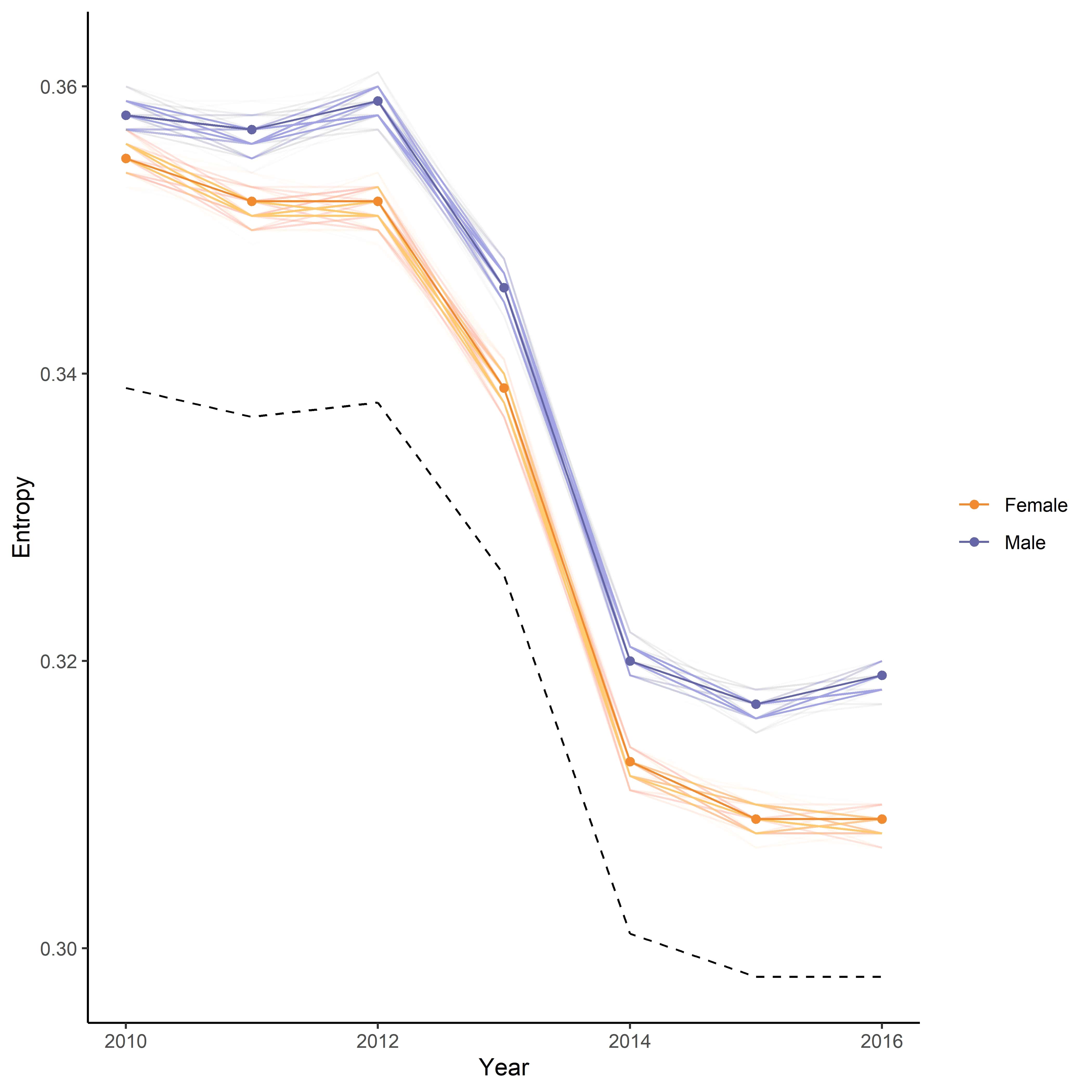}
    \caption{\textbf{Entropy by Sex Across Years}. Results show that entropy was higher for the male sub-population (purple) than the female sub-population (orange) and the baseline entropy of the population taken as a whole (dotted black). This indicates that enrolments for female students were more ordered, and male enrolments were more varied.
    }
    
    \label{fig:Entropy_Gender}
\end{figure}

The complete co-enrolment network across all years, regions, and deciles is shown in Figure \ref{fig:NetworkAll}). 
Across all years the infomap algorithm identified 42 communities of Level 3 STEM standards. As NCEA Level 3 is the most specialised stage of high school education, we would expect our network to be strongly partitioned into different community structures. This is reflected in a high modularity score of 0.83. The modularity score indicates that the nodes tend to share more edges with nodes within the same community than with nodes in different communities. The structure of the network changed over the period of time considered in the analysis, with a significant change taking place between 2012 and 2013. During this time, a change in education policy resulted in a reform in assessment. Science and mathematics linked unit standards were phased out, and a new set of achievement standards were introduced. Post the education reform in 2013, the overall number of standards diminished, and the network is mainly dominated by one community of mathematics and science standards (see Figure \ref{fig:NetworkYear}). This policy change is also reflected in changing levels of entropy in the network over time. As shown in Figure \ref{fig:Entropy_Gender}, the overall entropy of the network of assessments (taking all students into account) decreased over time. This gives an indication that, following the reforms to standards in 2013, student enrolments were more standardised and focused, and less flexible.

Through the use of network analysis we are able to delineate the main fields of study that comprise NCEA Level 3 STEM. Our method of using RCP and community detection separates out standards according to their propensity for being taken together, rather than simply classifying by subject label. The resulting network is partitioned according to two main pathways, communities of standards reflecting progression to university study (i.e., mainly achievement standards), and communities of standards orientated towards vocations (unit standards, and internally assessed standards). The detected communities thus provide a clearer picture of NCEA enrolment than broader subject labels. To provide an example, the chemistry standard ``\textit{Evaluate the interaction of a chemical process with society and/or the environment}'' may not assess the same content knowledge as another chemistry standard ``\textit{Demonstrate understanding of the properties of organic compounds}''. Despite both standards belonging to the chemistry  domain, the community detection algorithm assigned them to different communities in the network. While standards assessing applications of science to other vocations or to societal issues may help in the pathway to careers \textit{from} science, standards assessing scientific theory are more representative of the pathway to university and careers \textit{in} science.

On the whole, the communities in the network tended to be comprised of standards from the same domain of study. For example, biology standards tend to be taken in conjunction with other biology standards, physics with physics, and so on. However, the community detection algorithm mainly grouped science and mathematics subjects in the two large communities. These two communities, which occurred at different time periods (one before 2013, and one after) can be viewed as the pathway to university study. They consist mainly of achievement standards (many of them externally assessed, especially after 2013), and include physics, biology, chemistry, and calculus. 

The following sections will outline some patterns that can be observed from 2010 to 2016 by sex, ethnicity and school decile. While there are a vast number of patterns to be explored and discussed further, we focus our discussion on the main patterns. We provide the reader with full access to an interactive web application that can be used to explore the network in depth (\url{https://stur600.shinyapps.io/ExploreNCEA_L3_STEM/}). This application allows the user to filter the network by subject disciplines, types of standard, as well as school decile and demographic criteria. Through the patterns that we highlight, we seek to demonstrate the additional insights that can be gained through investigating the NCEA at a finer-grained level, and how they can further inform our understanding of what STEM participation looks like in Aotearoa New Zealand. We begin our discussion by focusing on the patterns that were seen based on students' sex, and then move on to discuss patterns by students' ethnicity and school decile.\\

\subsection*{Patterns by Sex}
Overall, there were small differences in the entropy in the network by sex, with entropy being slightly higher for the male sub-population (see Figure \ref{fig:Entropy_Gender}). This finding suggests that the male sub-population of the network had more enrolments spread across the network, while the female sub-population were more focused in specific areas. Further investigation of communities in the network showed clear examples of disparities in subject enrolments by sex which may explain the difference in entropy. Male students tended to dominate communities defined by standards in the agriculture, engineering, and practical technology (welding, furniture making etc.) domains, while female students had greater rates of enrolments in standards related to life sciences and textiles.

Corroborating the broad level trends outlined previously in the current study, and the trends detailed across other international contexts \citep{Else_Quest_2013, Sheldrake_2015, NSF, InstituteofPhysics_2013}, female students were more likely to enrol in biology standards, and less likely to enrol in physics and calculus standards. The majority of biology standards had around 60-70\% female students across years, while female students were also more likely to have enrolled in standards in the \textit{Core Science} domain. This domain includes standards such as \textit{Research a current scientific controversy} (61.5\% female) and \textit{Describe genetic processes} (67.3\% female). Female students were less likely to be represented in calculus and physics standards than male students. Investigating these disparities at the standard-level provides additional insights (see Table \ref{table:StdattainmentGender}). 

\begin{table}[htbp]
\begin{tabular}{|c|c|c|c|}
\hline
Standard & Assessment Type & Domain   & Female (\%) \\ \hline
     \parbox[c]{50mm}{\textit{Differentiate functions and use derivatives to solve problems}} & EX & Calculus & 38.2\\
      & & & \\
     \parbox[c]{50mm}{\textit{Integrate functions and use integrals to solve problems}} & EX & Calculus & 38.3\\
      & & & \\
     \parbox[c]{50mm}{\textit{Differentiate functions and use differentiation to solve problems}} & IN (Unit) & Calculus  &    42.3   \\
     & & & \\
     \parbox[c]{50mm}{\textit{Integrate functions and use integration to solve problems}} & IN (Unit) & Calculus & 46.3\\
      & & & \\
     \parbox[c]{50mm}{\textit{Demonstrate understanding of wave systems}} & EX & Physics  &  35.2\\
      & & & \\
     \parbox[c]{50mm}{\textit{Demonstrate understanding of electrical systems}} & EX & Physics  & 34.7 \\
      & & & \\
      \parbox[c]{50mm}{\textit{Demonstrate understanding of mechanical systems}} & EX & Physics & 36.2 
      \\ \hline
\end{tabular}  \caption{\textbf{Calculus and Physics Standard Enrolments by Sex.} Female students were underrepresented in calculus and physics standards in general, but especially in the external achievement standards that were part of the pathway to university science. The representation of female students in calculus unit standards was closer to even.}  \label{table:StdattainmentGender}

\end{table}

The rates of enrolment for female students in the physics standards were low, with the proportion of female students in externally assessed physics standards being around 35\% overall. The participation of female students in the standards related to calculus were also low compared to male students, with the proportion of female students being around 35-38\% in the main externally assessed standards. Interestingly, the internally assessed unit standard equivalents of the calculus standards, which were available to students prior to 2013, had an increased proportion of female students (around 42-46\%). Much research has been dedicated to understanding why disparities persist in physics and calculus by sex, with research often suggesting that female students tend to be less confident in mathematics and calculus compared to male students \citep{Hofer_2016, Heilbronner_2012, Simon_2015}. It may be that the calculus unit standards, which are assessed internally, in a familiar space with the opportunity to resit, offers a safer assessment environment where female students are more comfortable (see \citet{cheryan2017some} for a comprehensive review of the issues impacting on gender differences in STEM choice).\\

\subsection*{Patterns by Ethnicity and School Decile}
\begin{figure}[htbp]
    \centering
    \includegraphics[width = \textwidth]{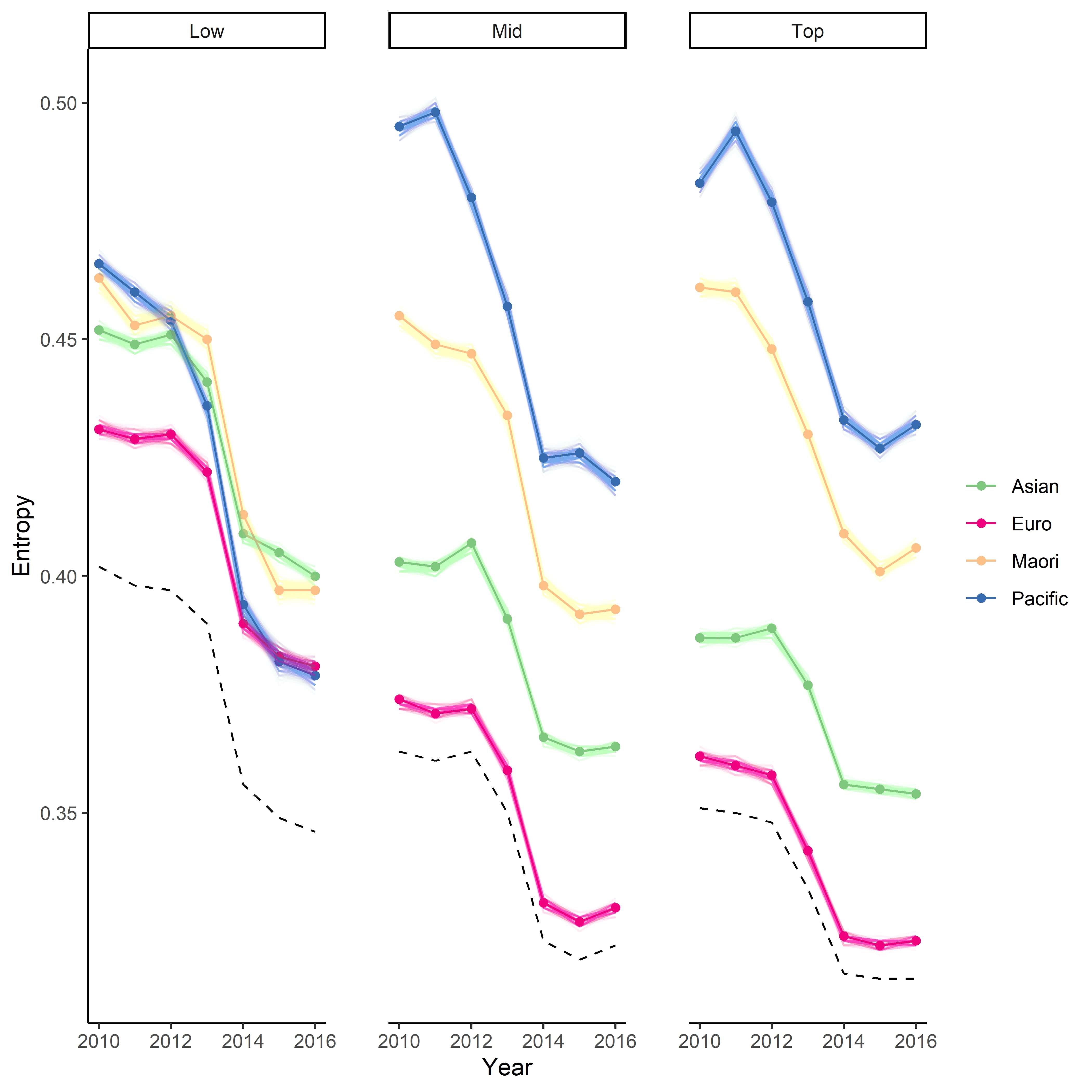}
    \caption{\textbf{Entropy by School Decile Across Years}. The entropy of each ethnic group sub-population, split by high school decile. The baseline entropy (black dotted line) is lower for the top decile schools (deciles 7-10), and higher for the low decile schools (deciles 1 to 3). While entropy is similar for ethnic groups in low decile schools, there are observable differences in entropy in middle and top decile schools, where the entropy is higher for M\={a}ori and Pacific sub-populations, and lower for Asian and European. This indicates that enrolments were more focused for students attending high decile schools, especially European students. 
    }
    
    \label{fig:Entropy_Decile}
\end{figure}

We report the results for ethnicity and school decile together, given that they are inextricably linked; M\={a}ori and Pacific students are over-represented in low decile schools. With the exception of students attending low decile schools, the M\={a}ori and Pacific sub-populations had higher entropy, while Asian and European students had lower entropy (see Figure \ref{fig:Entropy_Decile}).

\begin{figure}[htpb]
    \centering
    \includegraphics[width = \textwidth]{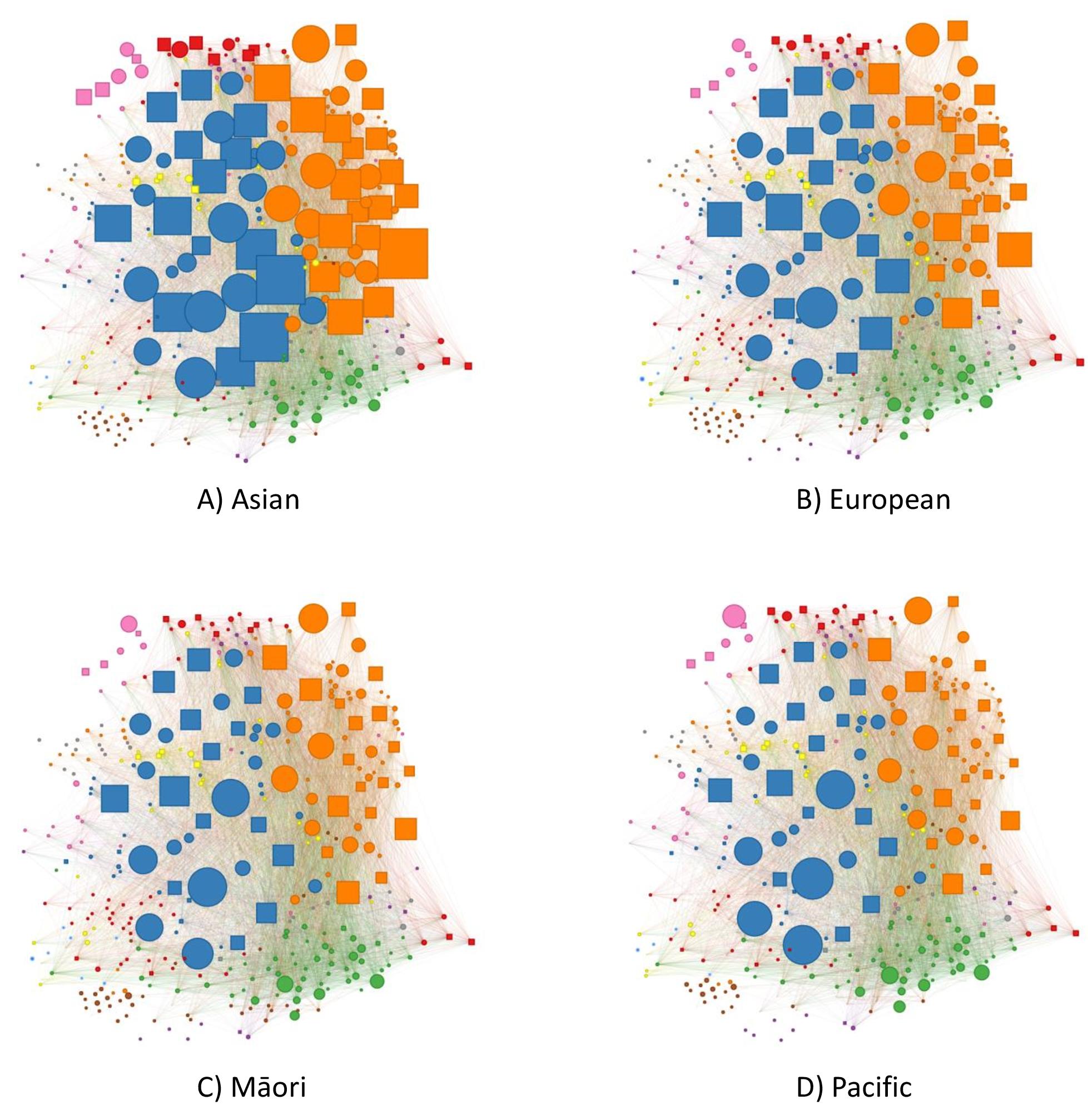}
    \caption{\textbf{Network of NCEA Level 3 Standards by Ethnicity}. The standard projection of the NCEA Level 3 standard co-enrolment network by ethnicity across all years. Node size represents the probability of a student from a sub-population being enrolled in a standard, colour represents community membership, squared nodes represent externally assessed standards, and circular nodes represent internally assessed standards. For all ethnic group sub-populations, the main science and mathematics communities (orange and blue nodes) tended to have higher probabilities of enrolment. The propensity for science and mathematics standard enrolment was especially true for Asian students, and less true M\={a}ori (C) and Pacific Island (D) groups.
    }
    
    \label{fig:NetworkEthnicity}
\end{figure}

As can be seen in Figure \ref{fig:NetworkEthnicity} the higher entropy for the M\={a}ori and Pacific sub-populations can be attributed to the fewer enrolments for M\={a}ori and Pacific students in science and mathematics (especially in the communities reflecting the pathways to university science), and the relatively increased enrolments in vocationally orientated standards. Asian and European students had more enrolments focused in science and mathematics, and were also over-represented in externally assessed standards, which are fewer in number than internally assessed standards. 

Although enrolments for the European and Asian sub-populations were focused in the communities of science and mathematics standards needed for university, the Asian sub-population had slightly higher entropy across years. The Asian sub-population had more enrolments in accounting standards, as well as mathematics and science standards overall, while the European sub-population had a narrower range of standards and relatively fewer enrolments in internally assessed science and mathematics standards. The higher entropy for Asian sub-population may relate to the categorical grouping of ``Asian'', which contains an extremely diverse population of students. This categorisation ranges from Pakistan and Bangladesh to China, and also some Pacific Island nations (e.g., Fijian Indians). The diversity of the population, including the cultures and social backgrounds, may have been reflected in an increased diversity of enrolments. 

Figure \ref{fig:Entropy_Decile} shows that entropy is more similar across ethnic groups for low decile schools, and that overall, the baseline entropy (indicated by the black line) for low decile schools is greater than the baseline entropy for higher decile schools. The higher entropy for low decile schools may be a consequence of increased enrolments in internally assessed standards, and fewer enrolments in standardised, externally assessed standards. Previous research has commented on this pattern \citep{hipkins2016ncea}. \citet{wilson2017subject} observed that lower decile schools were less likely to have students enrolled in Subject Literacy Achievement Standards, which are achievement standards that can be used as indicators of subject-specific literacy. After exploring the rates of enrolment, we also confirm that lower decile schools are less likely to have students enrolled in key externally assessed science and mathematics standards. 

While each ethnic group sub-population tends to have a similar entropy in low decile schools, a wider gap is present for middle and higher decile schools. For these schools, entropy is lower for Asian and European sub-populations, and higher for M\={a}ori and Pacific sub-populations. This suggests that the enrolment in STEM standards for M\={a}ori and Pacific students in middle and high deciles schools tends to be more diverse and less standardised. The lower entropy for Asian and European students is likely related to the focused participation in science and mathematics achievement standards required for university entrance, while enrolment for M\={a}ori and Pacific sub-populations is less focused in these standards and more balanced across other domains (including communities of unit standards). Table \ref{table:StdattainmentEthnicityDecile} shows the rates of enrolment in some key externally assessed science and calculus standards for each ethnic group sub-population, split by school decile and comparing 2010 to 2013.

\begin{landscape}
\begin{table}[htbp]
\begin{tabular}{lll|l|l|l|l|l|l|l|l|}
\cline{4-11}
          &&& \multicolumn{2}{l|}{\% of Asian} & \multicolumn{2}{l|}{\% of Euro} & \multicolumn{2}{l|}{\% of M\={a}ori}  & \multicolumn{2}{l|}{\% of Pacific} \\ \hline
\multicolumn{1}{|l|}{\parbox[c]{75mm}{\strut \centering Standard}} & \multicolumn{1}{l|}{\centering Domain}  &  \centering Year & Low & High & Low & High & \multicolumn{1}{l|}{Low} & High & Low & High \\ 
\hline
\multicolumn{1}{|l|}{\parbox[c]{75mm}{\strut \textit{Differentiate functions and use\\ derivatives to solve problems}}} & \multicolumn{1}{l|}{Calculus}  & 2010 & 36.2  & 52.2 & 17.7 & 23.1 & \multicolumn{1}{|l|}{11.3} & 15.9 & 13.4 & 15.2 \\
\hline
\multicolumn{1}{|l|}{\parbox[c]{75mm}{\strut \textit{Apply differentiation methods in solving\\ problems}}} & \multicolumn{1}{l|}{Calculus} & 2016 & 35.4 & 57.3 & 19.0 & 25.7 & \multicolumn{1}{l|}{9.9} & 16.3 & 13.6 & 18.3 \\ \hline
\multicolumn{1}{|l|}{\parbox[c]{75mm}{\strut \textit{Describe processes and patterns of\\ evolution}}} & \multicolumn{1}{l|}{Biology} & 2010 & 19.6 & 32.8  & 18.8 & 29.0 & \multicolumn{1}{l|}{10.9} & 23.3 & 7.4 & 17.0 \\
\hline
\multicolumn{1}{|l|}{\parbox[c]{75mm}{\strut \textit{Demonstrate understanding of evolutionary processes leading to speciation}}} & \multicolumn{1}{l|}{Biology} & 2016 & 27.2 & 33.7 & 24.0 & 31.8 & \multicolumn{1}{l|}{14.6} & 27.3 & 13.6  & 27.6  \\ \hline
\multicolumn{1}{|l|}{\parbox[c]{70mm}{\strut \textit{Describe aspects of organic chemistry}}} & \multicolumn{1}{l|}{Chemistry} & 2010 & 25.0 & 40.0 & 14.8 & 24.7 & \multicolumn{1}{l|}{7.4} &16.0 & 10.0 & 12.1 \\
\hline
\multicolumn{1}{|l|}{\parbox[c]{75mm}{\strut \textit{Demonstrate understanding of the\\ properties of organic compounds}}} & \multicolumn{1}{l|}{Chemistry} & 2016 & 32.0 &42.6 &19.9 &27.9 & \multicolumn{1}{l|}{13.0} & 19.6 &14.2 & 19.6 \\ 
\hline
\multicolumn{1}{|l|}{\parbox[c]{75mm}{\strut \textit{Demonstrate understanding of mechanical systems}}} & \multicolumn{1}{l|}{Physics} & 2010 & 25.4&38.0 &14.7 &23.9 & \multicolumn{1}{l|}{7.0} &16.7 & 8.4&11.0 \\
\hline
\multicolumn{1}{|l|}{\parbox[c]{75mm}{\strut \textit{Demonstrate understanding of mechanical systems}}} & \multicolumn{1}{l|}{Physics} & 2016 & 30.3&46.4 & 17.4& 27.3& \multicolumn{1}{l|}{8.5} &15.8 &9.6 &17.0 \\ \hline
\end{tabular}
\caption{\textbf{Key Standard Enrolments by Ethnicity and Decile}. The percentages of students enrolled in key externally assessed achievement standards in STEM by ethnic group and school decile (low/high). The percentage indicates the number of students from that ethnic group in a particular year who enrolled in the standard, as a fraction of the total number of students from that ethnic group in a particular year who took a STEM standard. For example, of the Asian students attending a low decile school in 2010 who took a STEM standard, 36.2\% took the calculus standard \textit{Differentiate functions and use derivatives to solve problems}. These percentages show that rates of enrolment differed across ethnic groups, with varied differences within these groups by school decile, and also comparing standards offered in 2010 and 2016.}  \label{table:StdattainmentEthnicityDecile}
\end{table}
\end{landscape}

While Table \ref{table:StdattainmentEthnicityDecile} shows that Asian and European had higher rates of participation in key externally assessed science and calculus standards, the differences between low and high school deciles appears to be considerable for the Asian sub-population compared to other groups. For example, the difference in participation for Asian students by decile in the calculus standard on differentiation offered in 2016 is 22\%, compared to 6.7\% for European, 6.4\% for M\={a}ori, and 4.7\% for Pacific Islands. This may once again point to the diversity of the categorisation of ``Asian'', but importantly highlights the importance of considering ethnic group categorisations in tandem with SES. 

The fact that low decile and M\={a}ori and Pacific sub-populations had fewer enrolments in key science and mathematics standards provides evidence that the pathway to university science is dominated by higher decile schools, and especially Asian and European students at these schools. In contrast, students from lower decile schools, and also M\={a}ori and Pacific students in higher decile schools, had relatively more enrolments in a larger and more disparate pool of internally assessed unit standards. Unit standards provide a valuable type of assessment that prepares students for vocational careers, and it may be that a higher proportion of students from less affluent areas seek vocational careers after high school. However, this does not necessarily explain why M\={a}ori and Pacific sub-populations attending higher decile schools are less likely to be channelled into science and mathematics standards. 

The differing patterns of enrolment for M\={a}ori and Pacific Island sub-populations and Asian and Pak\={e}ha is complex, but may be explained in several ways. Firstly, higher decile schools primarily serving M\={a}ori and Pacific students may choose to offer more internal assessments that provide increased opportunity to assess in culturally appropriate way (e.g., less competition, more formative feedback). Secondly, and less optimistically, it may be that teachers hold lower expectations for M\={a}ori and Pacific students \citep{turner2015teacher}, and are less likely to place them in the pathway towards university science. This idea was reflected by a participant in a study by \citep{graham2010Maori}: \blockquote{The teachers decide where the class is at in terms of choosing which standards [Unit versus Achievement]. It's a disadvantage on you because
it depends on what the teacher thinks you can do and what the kids in your class can do.}

Our analysis also highlights the impact of a pivotal reform in the NCEA, where curriculum-linked unit standards were phased out, and the system became more standardised and less flexible. Our results suggest that this change did not result in a decrease in participation in science and mathematics for M\={a}ori and Pacific Island sub-populations who were over-represented in the curriculum-linked unit standard in earlier years. Instead, as can be seen in Table \ref{table:StdattainmentEthnicityDecile} enrolment often increased at a greater rate than other ethnic groups, especially in higher decile schools. For example, in high decile schools the Pacific Island sub-population saw an larger increase of around 10.6\% in external biology standard relating to evolution, compared to 0.9\% for Asian, 2.8\% for European, and 4\% for M\={a}ori. Although we cannot comment on how this educational reform impacted on students' outcomes in science and mathematics, the reduced flexibility may actually help students by making NCEA less complex. Previous research has found that the complexity of NCEA can be confusing for students and parents to navigate \citep{graham2010Maori,jensen2010ncea}.\\

\subsection{Implications and Future Directions}
The current study fills a gap in the previous literature by investigating patterns of co-enrolments in NCEA Level 3 STEM standards by students' sex, ethnicity, and a proxy measure of SES. We believe that this study is the first of its kind to use bipartite networks to represent high school assessment data. Through our methodological approach, we are able to take into account a wealth of information related to students and the standards that they enrolled in. This includes demographic information (such as sex and ethnicity) and specific NCEA Level 3 standard information, such as the manner in which standards were assessed (externally or internally), and whether the standard was an achievement standard (traditional curriculum based subjects, such as English or science) or a unit standard (more vocational subjects, such as farming or practical technology).

The NCEA is very complex, but our method of analysis allows us to consider the different pathways that students follow based on the assessments they enrolled in. The communities of standards highlighted through our analysis reflect two main pathways, either towards vocations and careers \textit{from} science, or the pathway towards university and careers \textit{in} science. Despite growing discussion regarding the outcomes of different types of standards in the Aotearoa New Zealand context \citep{hipkins2016ncea,Lipson2017}, there has been a lack of research into how this information relates to student background. The methodology and results outlined in the current study enables us to represent the NCEA as a complex education system, and this can provide detailed insights into what science participation looks like.

A limitation of our analysis is the fact that we do not have access to students' level of achievement in the standards they enrolled in Level 3, or in previous years. As detailed by \citet{jensen2010ncea}, achievement outcomes in standards would be highly influential in shaping the pathways that open up or close off for students as they go through NCEA. Furthermore, the disparities seen in participation in key science standards may be tied to the development of academic identity \citep{bolstad2008seeing} which we are also unable to quantify. \citep[p.216]{Archer2014} argue that \blockquote{`cleverness' [can be viewed] as a racialized, gendered, and classed discourse, such that the identity of the `ideal' or `clever' student is not equally open to all students as a viable and authentic identity.} This notion of `cleverness' may explain the disparities found in the current study. More specifically, it may be that the `clever' pathway through NCEA is not open to all students. As described by \citet{hipkins2016ncea}, NCEA informally developed into a two-tiered system, with curriculum-linked unit standards commonly being viewed an easy pathway, and achievement standards, and especially externally assessed achievement standards, being viewed as a tougher pathway. Students who identify as less academic may purposefully seek easy pathways through NCEA, without fully understanding that doing so can reduce educational opportunities later on \citep{jensen2010ncea}.

Students with a family background of success in education may be more likely to view the academic pathway as normal or even expected. This idea is described in a related study of high school science pathways in the United Kingdom, where \citet{archer2017stratifying} found that students from more affluent backgrounds were more likely to see the science-orientated pathway as an `obvious' choice. Students from less affluent backgrounds may also be more motivated to seek full time employment, rather than pursue a pathway towards university study and the debt it may entail. However, the question remains as to the extent to which student from less affluent backgrounds knowingly choose vocational pathways and are not channeled down this pathway by simply attended a school in a low SES area.\\

\section{Conclusion}
The current study uses network science methods to explore disparities in science participation in Aotearoa New Zealand. It summarises the broad rates of participation by sex and school decile, and also participation at a finer-grained level through a network analysis of STEM standard co-enrolments for the final year of high school. The initial summary of science participation showed that male students have been more likely to take `physical' subjects (e.g., physics, calculus, practical technology), while female students have been more likely to take life science subjects (e.g., biology, health). A network analysis of NCEA enrolment data corroborated these findings, and added additional insights that showed that participation by sex were more equal in calculus unit standards. Our use of network analysis also allowed us to characterise the structure of co-enrolments for different sub-populations. Through the combination of Revealed Comparative Preference (RCP) and community detection, we were to explore the specific pathways that students participate in during high school STEM education, while a metric of entropy provided a description of how ordered or disordered co-enrolments were. This use of entropy to characterise co-enrolment provides a novel approach to understanding student pathways through education, and revealed valuable insights. We found that Asian and European sub-populations were had patterns of  enrolments focused in science and mathematics standards reflecting the pathway to university study. In contrast, the M\={a}ori and Pacific Island sub-populations, and lower decile school sub-population in general, had more disorganised patterns of enrolments with fewer enrolments in externally assessed science and mathematics standards. Our findings suggest that while policy changes have impacted on the structure of NCEA enrolments over time, disparities by sex, ethnicity, and school decile continued to be evident. While it is difficult to explain how much of standard enrolment is due to student choice, and how much of it is due to structural inequities present in the school system, our findings reveal disparities in STEM at a fine-grained level. Our findings suggest that the pathway to university science has been dominated by higher decile schools, and especially Asian and European students at these schools. These results provide a detailed picture of what STEM participation looks like in Aotearoa New Zealand.

\section{Disclaimer}
The results in this paper are not official statistics. They have been created for research purposes from the Integrated Data Infrastructure (IDI), managed by Statistics New Zealand. The opinions, findings, recommendations, and conclusions expressed in this paper are those of the author(s), not Statistics NZ. Access to the anonymised data used in this study was provided by Statistics NZ under the security and confidentiality provisions of the Statistics Act 1975. Only people authorised by the Statistics Act 1975 are allowed to see data about a particular person, household, business, or organisation, and the results in this paper have been confidentialised to protect these groups from identification and to keep their data safe. Careful consideration has been given to the privacy, security, and confidentiality issues associated with using administrative and survey data in the IDI. Further detail can be found in the Privacy impact assessment for the Integrated Data Infrastructure available from www.stats.govt.nz. 

\section*{Author Contributions}

SMT contributed to the conception, formulation, analysis, and writing of the manuscript. DO'N contributed to the formulation of the manuscript, and provided feedback.

\section*{Funding}
ST was supported by a University of Auckland Doctoral Scholarship (https://www.auckland.ac.nz/en.html). DO'N received funding from Te P\={u}naha Matatini (https://www.tepunahamatatini.ac.nz/) grant number UOA 9167-3705716. The funders had no role in study design, data collection and analysis, decision to publish, or preparation of the manuscript.

\section*{Acknowledgments}
We acknowledge the contribution of Adrian Ortiz-Cervantes who aided in the conceptualisation of this research.

\section*{Data Availability Statement}
The datasets generated for this study can be found at \url{https://stur600.shinyapps.io/ExploreNCEA_L3_STEM/} and R code is available on request.

\bibliographystyle{frontiersinSCNS_ENG_HUMS} 
\bibliography{main.bib}


\section*{Figure captions}

\end{document}